\documentclass[aps,pra,noshowpacs,twocolumn,preprintnumbers,superscriptaddress,floatfix]{revtex4}
\usepackage{graphicx}% Include figure files
\usepackage{dcolumn}% Align table columns on decimal point
\usepackage{bm}% bold math

\usepackage{times}

\usepackage{multirow}

\usepackage{epsfig,graphicx,amssymb} 

\usepackage{color}

\usepackage{latexsym}        
\usepackage{amsmath}             
\usepackage{amssymb}                 
\usepackage{amsbsy}
\usepackage{amsthm}
\usepackage{amsfonts}
\usepackage{mathrsfs}               
\usepackage{bm}                      
\usepackage{relsize}                 

\usepackage[T1]{fontenc} 
\begin{document}

%\preprint{APS/123-QED}

\title{Direct Observation of Broadband Nonclassical States in a Room-temperature\\
Light-matter Interface}

\author{Jian-Peng Dou}
\affiliation{State Key Laboratory of Advanced Optical Communication Systems and Networks, School of Physics and Astronomy, Shanghai Jiao Tong University, Shanghai 200240, China}
\affiliation{Synergetic Innovation Center of Quantum Information and Quantum Physics, University of Science and Technology of China, Hefei, Anhui 230026, China}
\author{Ai-Lin Yang}
\affiliation{State Key Laboratory of Advanced Optical Communication Systems and Networks, School of Physics and Astronomy, Shanghai Jiao Tong University, Shanghai 200240, China}
\affiliation{Synergetic Innovation Center of Quantum Information and Quantum Physics, University of Science and Technology of China, Hefei, Anhui 230026, China}
\author{Mu-Yan Du}
\affiliation{State Key Laboratory of Advanced Optical Communication Systems and Networks, School of Physics and Astronomy, Shanghai Jiao Tong University, Shanghai 200240, China}
\author{Di Lao}
\affiliation{State Key Laboratory of Advanced Optical Communication Systems and Networks, School of Physics and Astronomy, Shanghai Jiao Tong University, Shanghai 200240, China}
\author{Hang Li}
\affiliation{State Key Laboratory of Advanced Optical Communication Systems and Networks, School of Physics and Astronomy, Shanghai Jiao Tong University, Shanghai 200240, China}
\affiliation{Synergetic Innovation Center of Quantum Information and Quantum Physics, University of Science and Technology of China, Hefei, Anhui 230026, China}
\author{Xiao-Ling Pang}
\affiliation{State Key Laboratory of Advanced Optical Communication Systems and Networks, School of Physics and Astronomy, Shanghai Jiao Tong University, Shanghai 200240, China}
\affiliation{Synergetic Innovation Center of Quantum Information and Quantum Physics, University of Science and Technology of China, Hefei, Anhui 230026, China}
\author{Jun Gao}
\affiliation{State Key Laboratory of Advanced Optical Communication Systems and Networks, School of Physics and Astronomy, Shanghai Jiao Tong University, Shanghai 200240, China}
\affiliation{Synergetic Innovation Center of Quantum Information and Quantum Physics, University of Science and Technology of China, Hefei, Anhui 230026, China}
\author{Lu-Feng Qiao}
\affiliation{State Key Laboratory of Advanced Optical Communication Systems and Networks, School of Physics and Astronomy, Shanghai Jiao Tong University, Shanghai 200240, China}
\affiliation{Synergetic Innovation Center of Quantum Information and Quantum Physics, University of Science and Technology of China, Hefei, Anhui 230026, China}
\author{Hao Tang}
\affiliation{State Key Laboratory of Advanced Optical Communication Systems and Networks, School of Physics and Astronomy, Shanghai Jiao Tong University, Shanghai 200240, China}
\affiliation{Synergetic Innovation Center of Quantum Information and Quantum Physics, University of Science and Technology of China, Hefei, Anhui 230026, China}
\author{Xian-Min Jin}
\thanks{xianmin.jin@sjtu.edu.cn}
\affiliation{State Key Laboratory of Advanced Optical Communication Systems and Networks, School of Physics and Astronomy, Shanghai Jiao Tong University, Shanghai 200240, China}
\affiliation{Synergetic Innovation Center of Quantum Information and Quantum Physics, University of Science and Technology of China, Hefei, Anhui 230026, China}
%\date{\today}

\pacs{Valid PACS appear here}% PACS, the Physics and Astronomy

\maketitle

\textbf{Nonclassical state is an essential resource for quantum-enhanced communication, computing and metrology to outperform their classical counterpart. The nonclassical states that can operate at high bandwidth and room temperature while being compatible with quantum memory are highly desirable to enable the scalability of quantum technologies. Here, we present a direct observation of broadband nonclasscal states in a room-temperature light-matter interface, where the atoms can also be controlled to store and interfere with photons. With a single coupling pulse and far off-resonance configuration, we are able to induce a multi-field interference between light and atoms to create the desired nonclassical states by spectrally selecting the two correlated photons out of seven possible emissions. We explicitly confirm the nonclassicality by observing a cross correlation up to 17 and a violation of Cauchy-Schwarz inequality with 568 standard deviations. Our results demonstrate the potential of a state-built-in, broadband and room-temperature light-matter interface for scalable quantum information networks.}

\section*{INTRODUCTION}
Quantum information network, large enough to be capable of carrying out quantum technologies \cite{OBrien2009,Gisin2007,Ladd2010,Guzik2012} beyond classically possible, is composed of two key elements: the nonclassical state carrying flying qubit and quantum memory as the node of storing stationary qubit. An efficient architecture requires both nonclassical state and quantum memory to have high fidelities, mutually compatible light-matter interfaces, and preferably to have the capabilities of running at high speed and room temperature conditions \cite{Lvovsky2009,Sangouard2011}. 

\begin{figure*}
\centering
\includegraphics[width=1.5\columnwidth]{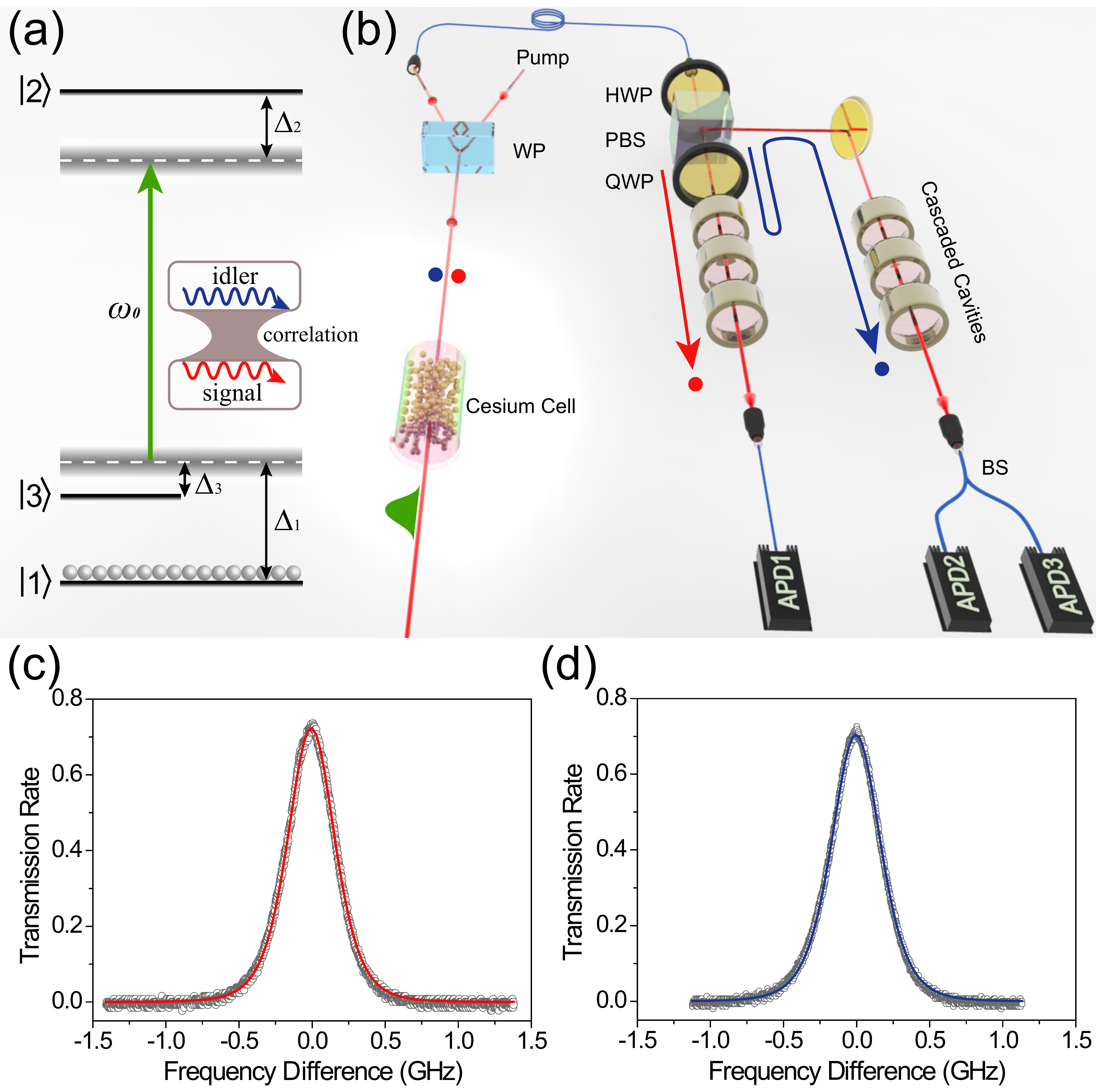}
\caption{\textbf{Schematic view of the experiment.} \textbf{(a)} Energy-level diagram. $\omega_0$ is the central angle frequency of the coupling pulse. $\Delta_1$, $\Delta_2$, $\Delta_3$ are the detunnings with positive values. \textbf{(b)} Experimental setup. The cesium cell is a 75\,mm-long cylindrical glass filled with $^{\rm 133}$Cs atoms and 10\,Torr neon as buffer gas, and its temperature is heated up to 61.3\,$^{\circ}$C. WP, Wollaston prism; HWP, half-wave plate; QWP, quarter-wave plate, PBS, polarization beam splitter; APD, avalanche photodiode. Transmission properties of the cascaded cavities built for signal \textbf{(c)} and idler \textbf{(d)} photons. The total full width at half maximum (FWHM) of each set of cascaded cavities is around 380\,MHz, and their peak transmission efficiencies exceed 70\%. } \label{Figure1}
\end{figure*}

Albeit there are advances of pushing bandwidth from narrowband to broadband and storage media from ultra-cold atomic gas to room-temperature atomic vapour \cite{Chaneliere2005,Eisaman2005,Zhang2011,Kuzmich2003,Chrapkiewicz2017,Julsgaard2004,Alexander2006,Afzelius2009,Hosseini2011,Reim2011,Ding2015}, it is not until recently that the room-temperature broadband memories being capable of operating with a high fidelity genuinely in quantum regime were achieved \cite{Kaczmarek2018,Dou2017}. However, the spectrum of conventional optical down conversion by pumping a nonlinear crystal is still too large to match the acceptance bandwidth of quantum memories. Although a cavity can be added to shape the spectra and enhance the efficiencies of down converted photons, it is extremely challenging to align and maintain the nonclassical state generation highly efficient and stable \cite{Ou1999,Kuklewicz2006,Zhang2011}. Moreover, the specially-built state generation systems and their complexities make quantum networks physically hard to scale up. 

In case of that the nonclassical state can be directly obtained by employing the same atomic ensemble and similar light-matter interaction mechanism with quantum memory, the resulting naturally-matched emission line and bandwidth will make quantum networks elegant and straightforward to build. In the last 15 years, many works have successfully demonstrated nonclassical state (heralded single photons or photon pair sources) generations in cold atoms \cite{Kuzmich2003,Chou2004,Thompson2006,SNarrow02,Shu2016}. However, the room-temperature and broadband features are more appealing and essential for being compatible with high-speed and physically scalable quantum technologies, and meanwhile have been proven very challenging \cite{van der Wal2003,Manz2007,Bashkansky2012}. The previous experiments with small detuning in room-temperature atoms  suffer from fluorescence noise induced by Doppler effect and atom-atom collision. The broadband nonclassical correlation can be observed by continuously addressing ladder-type hot atoms, which is, however, not directly usable due to the random creation time of correlated photons and the distinctly different interaction mechanism with Raman process of broadband quantum memory\cite{Willis2010,Ding2012}.

In this work, we present a multi-field light-matter interference induced by a far-off resonance short coupling pulse. We have addressed all these open questions by realizing nonclassical discrete-variable states generation and measurement at pulse mode, broad bandwidth and room temperature. Furthermore, the  collinear configuration realized in our experiment may be used for long lifetime quantum memory due to minimization of spin wave dephasing.

\begin{figure*}
\centering
\includegraphics[width=1.6\columnwidth]{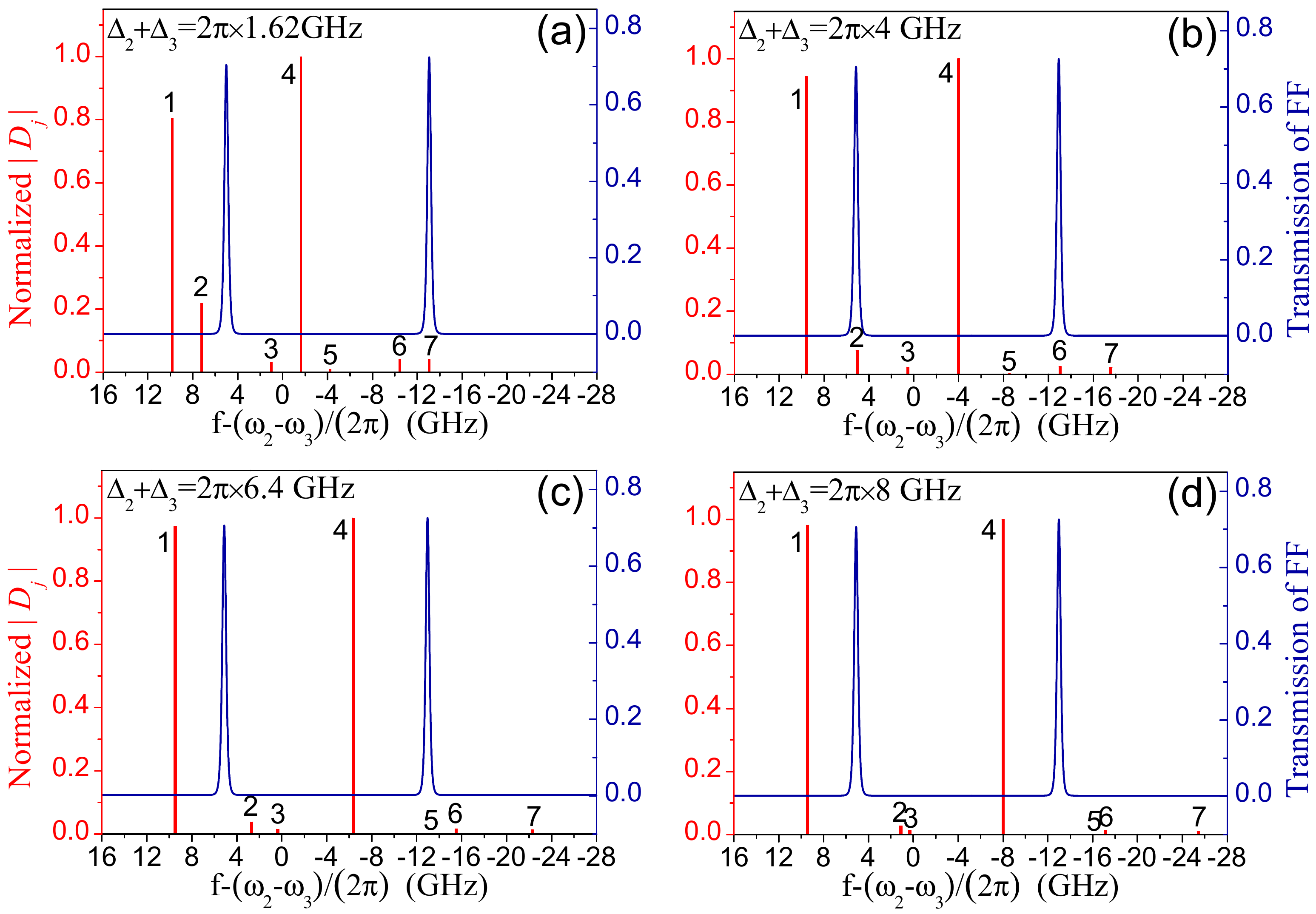}
\caption{\textbf{Emissions of the multi-field interference induced by a single coupling pulse.} The normalized components $\left | D_j \right |$ with four different detunings of the coupling light and their Y-axis labels are marked in red. The transmission windows of the frequency filter (FF) constituted by cascaded cavities and their Y-axis labels are marked in blue. They are shown together in order to facilitate the selection of emissions and explanation of the observed three-peak phenomena.}
\label{Figure2}
\end{figure*}

\begin{figure}
\centering
\includegraphics[width=1\columnwidth]{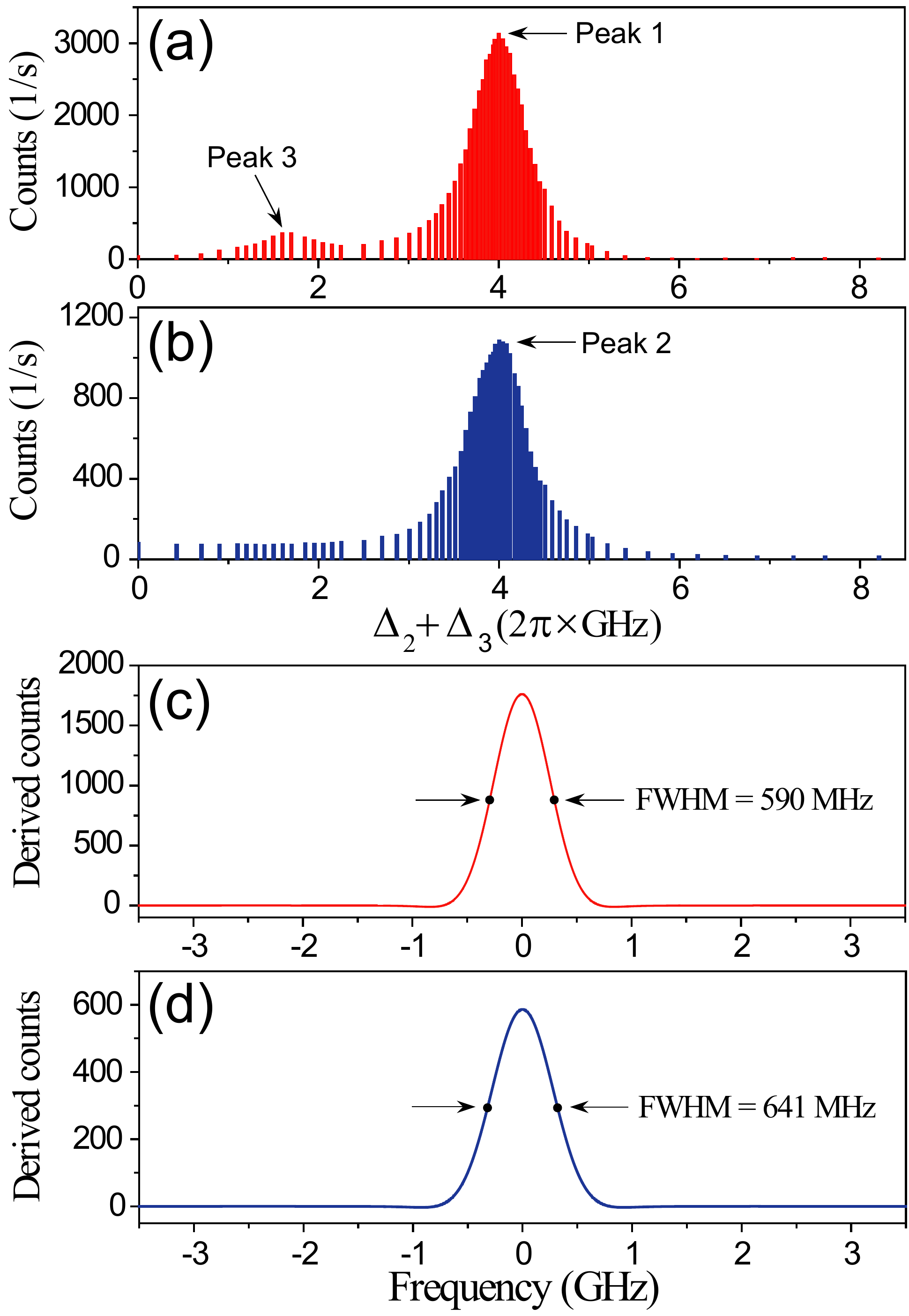}
\caption{\textbf{Experimental results of convolution measurement on the correlated photons.} The measured convolution of the signal \textbf{(a)} and idler \textbf{(b)} photons and the windows of cascaded cavities. Peak 1 and Peak 2 correspond to signal and idler photons respectively. Peak 3 is explained in main text. From the spectra of signal \textbf{(c)} and idler \textbf{(d)} photons, their bandwidths can be derived as 590 MHz and 641 MHz respectively.} \label{Figure3}
\end{figure}

\section*{RESULTS}
The far off-resonance and $\Lambda$-type configuration shown in Fig.\,1(a) are the two main features that our scheme shares with Raman \cite{Reim2011} and FORD \cite{Dou2017} broadband quantum memory. The two lower states $\left | 1 \right \rangle \left( 6S_{1/2}, F=3\right)$ and $\left | 3 \right \rangle\left( 6S_{1/2}, F=4\right)$ are the hyperfine ground states of $^{\rm 133}$Cs with a frequency difference of 9.19\,GHz, and the upper state $\left | 2 \right \rangle$ is the excited state that we mark at the cross over between $6P_{3/2}, F'=4 $ and $F'=5$ as a precise reference. We initialize all atoms to the state $\left | 1 \right \rangle$ by a pump light in the transition of $\left | 3 \right \rangle$ and $\left | 2 \right \rangle$, and temporally switch off the pump by an acoustic optical modulator. We apply a single linearly polarized and 2\,ns short coupling pulse with a detuning of $\Delta_2+\Delta_3$. In 1969, Mollow found that a single spectral line of a two-level atom driven near resonance by a monochromatic classical electric field can develop to three spectral lines called Mollow-triplet emission spectrum\cite{Mollow1969}. The investigation later on suggested that three-level atoms possess richer interference effects. The Mollow-triplet sideband emission from a quantum dot has been used to generate cascaded single photons\cite{Ulhaq2012}. It is reasonable to consider the light-matter interference in Raman-like system as another counterpart of the Mollow-triplet. The interaction can be regarded as a straightforward model of a light-matter multi-field interference driven by single far off-resonance pulse (see Methods). We can derive the frequency position and relative emission amplitude  $\left | D_j \right |$ of each spectral component. 

In Fig.\,2, we show the normalized amplitude $\left | D_j \right |$ with four different detunings of the coupling light. Compared with two-level atoms, three-level atoms possess richer interference effects. During the interaction, the strong coupling pulse can induce some virtual energy levels, which is the reason why some extra spectral components appear, for example the component 7 in Fig.\,2. The concept of virtual energy level is also used to illustrate the Raman scattering. We select correlated components 2 and 6 as the idler and signal respectively for the observation of nonlcassical states. The component 1 corresponds to the resonant transition $\left| {\rm{2}} \right\rangle  \to \left| {\rm{1}} \right\rangle $. In addition, most of atoms populate in the state $\left| {\rm{1}} \right\rangle $ (see Fig.\,5 in the Methods). There is a strong coupling between the state $\left| {\rm{2}} \right\rangle $ and state $\left| {\rm{1}} \right\rangle $. Thus, the component 1 is consistently large. In contrast, very few atoms populate in the state $\left| {\rm{3}} \right\rangle $, and hence the component 3 is much smaller, although it corresponds to the resonant transition $\left| {\rm{2}} \right\rangle  \to \left| {\rm{3}} \right\rangle $. It means that the amplitude of spectral components depends not only on the detuning but also the population. In our experiment, we did not choose the component 1 and 3 because they are resonant or near-resonant to the atomic transition line, and therefore inevitably suffer from contamination of fluorescence noises\cite{Hsu2006}. The component 4 overlaps with the coupling light while component 5 is considerablly small. Both the theoretical and experimental results confirm that $\left | D_5 \right |$ is vanishingly small. This suggests that the amplitude of the corresponding virtual energy level is much smaller. The detected counts of the component 7 are also much smaller than components 2 and 6 (see Fig.\,3(a)). With Fig.\,2, we can also determine that the optimal detuning is around 4\,GHz when both signal and idler are far away from the resonant fluorescence. 

It will be much more complex to calculate the emission direction and polarization characteristics of each spectral component. In the light of two previous works \cite{Duan2002,Nunn2008}, we use a high-extinction-ratio Wollaston prism and a single mode fiber to well define the desired polarization and emission direction. The emission direction of the desired correlated photons is coaxial with the coupling light, and their polarizations are perpendicular to that of the coupling light. 

As is shown in Fig.\,1, by independently setting the central frequency position of two sets of cascaded cavities, we clearly observe the desired idler (blue dot) and signal (red dot) when the detuning $\Delta_2+\Delta_3= 2\pi\times4$ \,GHz. Then we scan the detuning $\Delta_2+\Delta_3$ while the cascaded cavities remain unchanged, and we can obtain the convolution between the window of cascaded cavities and the desired photons (see Fig.\,3(a) and (b)). The Peak 1 and Peak 2 correspond to the signal and idler photons respectively. 

\begin{figure}
\centering
\includegraphics[width=0.95\columnwidth]{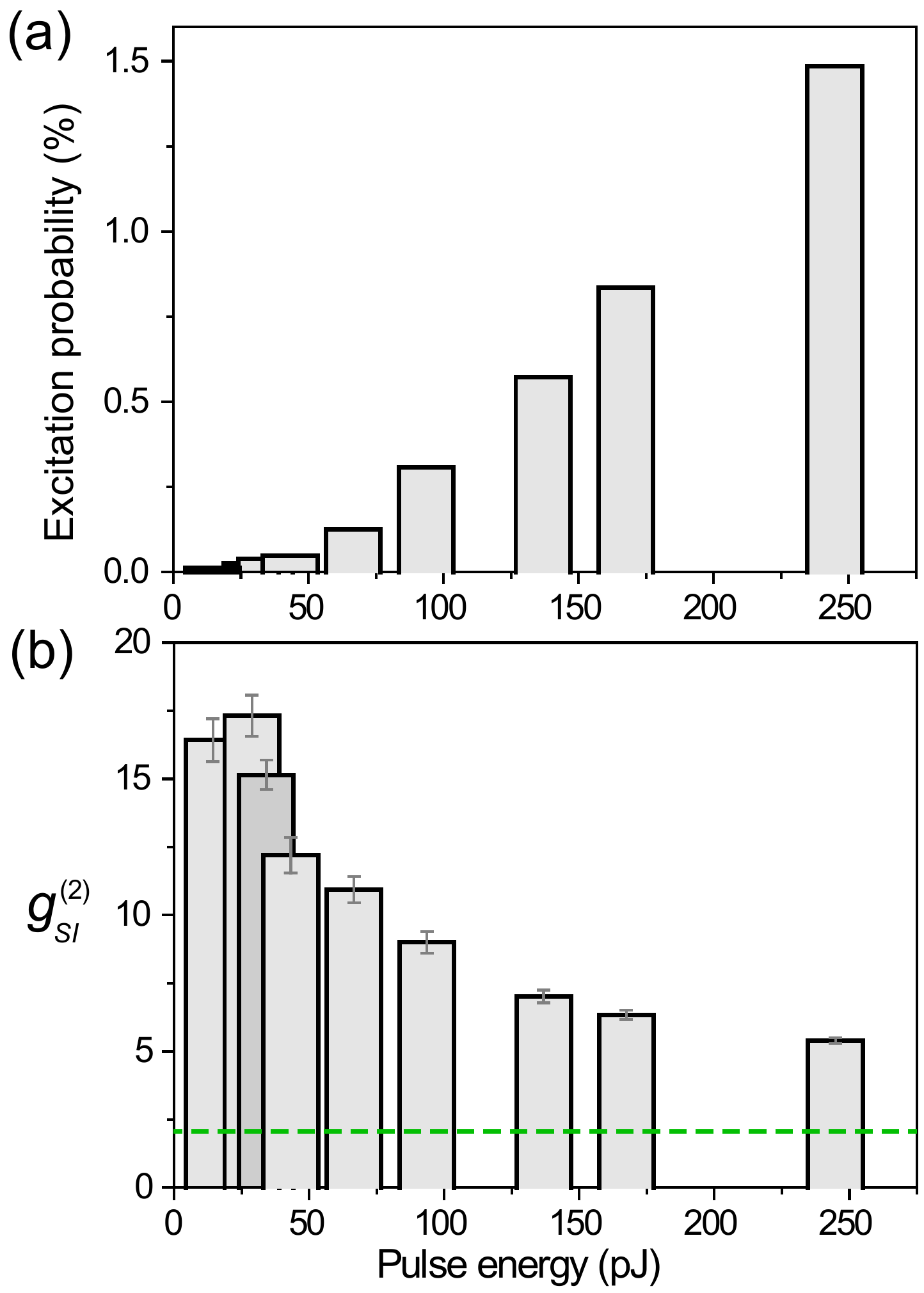}
\caption{\textbf{Nonclassicality-excitation tradeoff.} Experimental results on excitation probability of photon pairs \textbf{(a)} and cross correlation function $g_{SI}^{(2)}$ \textbf{(b)} with scanned pulse energy of the coupling light. The detuning is set at $\Delta_1+\Delta_2=2\pi\times4$\,GHz. The results $g_{SI}^{(2)}$ larger than 2 (green dashed) is in nonclassical regime. The beam waist of the coupling light is 90\,$\mu {\rm m}$. The coincidence window (also known as integration gate) is 9 ns and is centered at the arrival time of the desired photons. Error bars are given by Poissonian statistics.
} \label{Fig4}
\end{figure}

In addition to the expected peaks, the appearance of Peak 3 needs to be explained. As we can see in Fig.\,2(a), when $\Delta_2 +\Delta_3 =2\pi\times1.62$\,GHz, the component 7 apparently overlaps with one of the transmission windows of the cascaded cavities, which results in the emergence of Peak 3 shown in Fig.\,3(a). The measured component 7 is much weaker than the component 6, which implies that the polarization or emission direction of component 7 is much different from the desired direction defined by the Wollaston prism and the single mode fiber. The idler and signal photons correspond to the components 2 and 6 respectively, and are designed to maximally pass through cascaded cavities when $\Delta_2 +\Delta_3 =2\pi\times4.00$\,GHz. For other detunings such as $2\pi\times6.40$\,GHz and $2\pi\times8.00$\,GHz, there are no any other overlap with the transmission windows of the cascaded cavities. The three peaks that we have observed therefore further confirm our theoretical model. 

The bandwidth of the generated photons can be mainly determined by the bandwidth of coupling pulses, Doppler broadening, collision broadening and AC-Stark effect. In our experiment, while the larger bandwidth of nonclassical states can be achieved by using shorter coupling pulses, it should be smaller than cesium ground state hyperfine splitting of 9.19 GHz. With the measured transmission window of the cascaded cavities shown in Fig.\,1(c), (d) and their convolution with the desired photons shown in Fig.\,3(a), (b), we obtain the spectra of the signal and idler photons shown in Fig.\,3(c), (d) by applying convolution theorem and Fourier transform. The bandwidths of the signal and idler photons are 590\,MHz and 641\,MHz respectively, which confirms the broadband property of the correlated photons that are determined by the pulse duration of the coupling light (2\,ns). This observed broadband nonclassical state, associated with a well-defined creation time, is therefore well compatible with Raman \cite{Reim2011} and FORD \cite{Dou2017} broadband quantum memory for future quantum enhanced applications.

To quantify the nonclassicality of the correlated photons, when the detuning $\Delta_2+\Delta_3=2\pi\times4$\,GHz and the coupling pulse energy 95\,pJ, we measure the violation of Cauchy-Schwarz inequality\cite{Clauser1974}
\begin{equation} 
\left( g_{SI}^{(2)} \right)^2 \leq g_{SS}^{(2)}\,\cdot \, g_{II}^{(2)}
\end{equation}
where correlation functions can be obtained from the detection probability, for example,  the cross correlation $g_{SI}^{(2)}=p_{SI}/(p_{S} \cdot p_{I})$. The non-heralded auto-correlation of the signal (idler) photons $ g_{SS}^{(2)}$ ($g_{II}^{(2)}$) is $2.05 \pm 0.10$ ($1.64 \pm 0.21$), and the cross correlation of the signal and idler photons $g_{SI}^{(2)} $ is up to $8.58 \pm 0.12$. Our measurements lead to a violation of Cauchy-Schwarz inequality up to 568 standard deviations.

In an ideal condition, $g_{SS}^{(2)} =2 $ and  $g_{II}^{(2)} =2 $. In practice, unwanted multimode couplings and the background noises (such as the leakage of the coupling pulses and dark counts) may decrease the auto-correlation, thus $g_{SS}^{(2)} $ and $g_{II}^{(2)} $ may go a bit smaller than 2. Although there always exists a fluctuation around the ideal value due to the instability of the experimental setup, we still find that $g_{SS}^{(2)}=2.05 \pm 0.10$ is almost equal to 2 with a measurement uncertainty $\pm 0.1$ based on Poissonian statistics, which means that the noise in the signal channel can be ignored and means a high heralded purity\cite{Spring2013}. The cross correlation $g_{SI}^{(2)} > 2$ is a signature of the violation of the Cauchy-Schwartz inequality\cite{Sangouard2011}. Compared with classical light, the essential feature of non-classical light is that the non-classical photons are correlated with each other, i.e. the detection of the signal photon can herald the existence of the idler photon. The heralded single photon itself is a crucial source for many quantum tasks. With the help of quantum memory, many heralded single photons can be effectively synchronized to multiphoton sources for large-scale quantum information processing. In contrast, the detection of a classical photon, such as a chaotic light or laser, cannot herald the existence of the other photon. That is to say whether or not the other photon comes is completely uncertain, which invalidates the application of quantum memories in photonic networks. Due to the importance of the Cauchy-Schwartz inequality, it is widely used in the field of quantum information \cite{Chaneliere2005, Eisaman2005, Zhang2011}. 

In Fig.\,4, the measured excitation probability of photon pairs and cross correlation $g_{SI}^{(2)} $ show a tradeoff as a function of the coupling pulse energy. The coupling pulse width is 2ns, and the width of the arrival time of the generated photons is also about 2ns. In our experiment, the coincidence window (also known as integration gate) is 9 ns and is centered at the arrival time of the desired photons. If the counting system obtains a signal photon and an idler photon in a same trial (i.e., the two photons are generated by a same coupling pulse), the counting system will record the event which indicates a successful generation of photon pairs. By accumulating a lot of trials, we can obtain the mean number of photon pairs per coupling pulse. The results shown in Fig.\,4(a) have been divided by the square of the total detection efficiency and show the excitation probability originally in the atomic ensemble. The maximum total transmission efficiency of the cascaded cavity filters is about 70\% as is shown in Fig.\,1 (c) and (d). Based on the spectra of signal and idler photons shown in Fig.\,3(c) and (d), we can calculate out the practical transmission efficiency about 40\%. Taking collection efficiency of fibers into account, the total transmission efficiency is about 11\%. The quantum efficiency of detectors is about 50\%. Due to the symmetry of the experimental setup, the total detection efficiency of the signal photons is almost same to that of the idler photons. The total detection efficiency is about 5.5\%.

We can see that the cross correlation $g_{SI}^{(2)} $ can reach 17 at a lower excitation probability (see Fig.\,4(b)). For a comparably large excitation probability, $g_{SI}^{(2)} $ all exceed the classical boundary of 2. There is a probability that the emitted photons are re-absorbed, which is called radiation trapping \cite{Thomas2017}. It is a common phenomenon in various light-matter interaction processes. However, to obtain high-quality nonclassical states, we only need to care about the effective excitation probability. We also experimentally investigate the time dependence properties of cross correlation (see Methods).

For a photon pair, $p_{SI}/p_{S}$ is the heralding efficiency for idler photons. In our experiment, the heralding efficiency changes from 1.5\% to 16\% as the excitation probability (photon pairs per coupling pulse) changes from 0.013\% to 1.49\%. The probability $p_I$ of idler photons changes from 0.09\% to 3.0\%. We can see that the growth rate of the probability $p_I$ is higher than that of the heralding efficiency, therefore results in a decrease of the cross correlation shown in Fig.\,4(b). When the idler photons is used as the herald, then the heralding efficiency is given by $p_{SI}/p_{I}$ which changes from 14\% to 50\% as the excitation probability changes from 0.013\% to 1.49\%. Thus, we can get a considerably larger heralding efficiency when the idler photons is the herald.

It is worth mentioning that the pump light used to prepare the atoms in initial state $\left | 1 \right \rangle$ is necessary for achieving the nonclassicality. Without the pump light, the cross correlation $g_{SI}^{(2)}$ degrades to $1.62\pm 0.04$, i.e., the nonclassicality disappears. If the ground state is unpumped, atoms almost equally populate in state $\left | 1 \right \rangle$ and state $\left | 3 \right \rangle$. A coupling pulse does not only excite photons from state $\left | 1 \right \rangle$ but also from state $\left | 3 \right \rangle$, and the two processes are independent to each other, i.e. the photons from the state $\left | 1 \right \rangle$ have no correlation with that from the state $\left | 3 \right \rangle$. Thus, the non-classicality disappears when the ground state is unpumped. In an ideal condition, the purity of the initial state should be 100\%, i.e., all of the atoms populate on the energy level $\left | 1 \right \rangle$ after the pumping process, which is challenging for warm atoms. We use a pump light with a power of 16.4 mW and a pump time of 1.2 us. The beam waist of the pump light is about 2500 um. We use a very weak and counter-propagated probe pulse (pulse duration 300 ns, beam waist 140 um) to measure the polarization purity. Since the pumped region is one order larger than the probe region, and the time delay of 300\,ns between the probe pulse and the turnoff of the pumping light is so short that the unpumped atoms outside the pumped region can not enter the probe region. We finally achieve a purity of the initial state up to 90\%.

\section*{DISCUSSION}
In summary, we propose and experimentally demonstrate a multi-field interference between light and atoms with single coupling pulse and far off-resonance configuration. With this we achieve the direct observation of broadband nonclasscal states in a room-temperature light-matter interface, where the atoms can also be controlled to store and interfere with photons. The obtained nonclassical states are compatible and therefore can be directly used to interfere with broadband off-resonance quantum memories\cite{Reim2011,Dou2017}.

To be compatible with large-scale fiber networks, the near-infrared wavelength around 852\,nm of the obtained nonclassical state can be converted to telecommunication wavelength by using the cascade schemes\cite{Radnaev2010,Dudin2010} or on-chip frequency converters \cite{Albrecht2014,Farrera2016,Ikuta2016}.

The measured strong nonclasicality associated with low noise level reveals the potential of the collinear configuration, in which the created photons are coaxial with the coupling light, for both broadband state generations and quantum memories in room-temperature atoms. This is particularly important for the architectures based on DLCZ protocols\cite{Kuzmich2003,Chrapkiewicz2017,Duan2001,Duan2002,Bashkansky2012,Namazi2017} as the emitted photons correlated with the spin waves stored in the quantum memory are mainly inside a small cone around the direction of the write light\cite{Duan2002}. Hence, a collinear configuration means the highest collection efficiency of write-out photons. Additionally, a collinear configuration possesses a maximum wavelength of spin waves and therefore a longer storage lifetime against dephasing\cite{Zhao2009}.

\section*{Acknowlegements}
The authors thank J.-W. Pan for helpful discussions. This work was supported by National Key R\&D Program of China (2017YFA0303700); National Natural Science Foundation of China (NSFC) (11374211, 61734005, 11690033); Shanghai Municipal
Education Commission (SMEC)(16SG09, 2017-01-07-00-02-E00049); Science and Technology Commission of Shanghai Municipality (STCSM) (15QA1402200, 16JC1400405); Open fund from HPCL (201511-01). X.-M.J. acknowledges support
from the National Young 1000 Talents Plan.\\

\section*{Methods}
\subsection{\textbf {A light-matter multi-field interference driven by single far off-resonance pulse:}}
The total Hamiltonian of atoms can be written as
\begin{equation}
H=H_0+H_{\rm I}
\end{equation}
where $H_0\left | n \right \rangle=\hbar \omega_n \left |n\right \rangle$ with $n=1, 2, 3$ and interaction term is $H_{\rm{I}}=e\vec R \cdot \vec E $. The vector $\vec R$ is the position of the outermost electron of Cs, and $\vec E=\frac{1}{2}\vec A\left(\vec r, t\right)\left[\exp \left( -i\omega_0t \right)+ \exp \left(i\omega_0t \right)\right]$ denotes the coupling light. The wave function $\psi \left( \vec r, t\right)$ of atoms affected by coupling light can be expanded in terms of the eigenfunction $\left | n \right \rangle$ of bare atoms.
\begin{equation}
\psi \left( \vec r, t\right)= \sum\limits_{n = 1}^3 {\left \{ C_n (t)\left| n \right\rangle \exp \left[- i \omega _n t+ (-1)^n i\Delta_n  t \right]\right \}}
\end{equation}

By using rotating-wave approximation, the Schr\"{o}dinger equation $i\hbar \frac{\partial \psi \left( \vec r, t\right)}{\partial t}=H\psi \left( \vec r, t\right)$ can be expressed as a form of matrices
\begin{equation}\label{juzhenH}
i\hbar \frac{\partial}{\partial t} \left(
{\begin{array}{*{20}c}
{C_1}\\
{C_2}\\
{C_3}\\
\end{array}}
\right)=
\frac{\hbar }{2}
\left(
{\begin{array}{*{20}c}
   { - 2\Delta _1 } & {\Omega _{12} } & {0}  \\
   {\Omega _{21} } & {2\Delta _2 } & {\Omega _{23} }  \\
   {0} & {\Omega _{32} } & { - 2\Delta _3 }  \\
\end{array}}
\right)
\left( {\begin{array}{*{20}c}
{C_1}\\
{C_2}\\
{C_3}\\
\end{array}}
\right)
\end{equation}
where $\Omega_{mn} = { \left \langle m \left | e \vec R \cdot \vec A  \right | n \right \rangle}/{\hbar}$. For simplicity, the space-time dependent $\Omega_{mn}$ can be replaced by its effective value.

\begin{figure*}
\centering
\includegraphics[width=0.95\columnwidth]{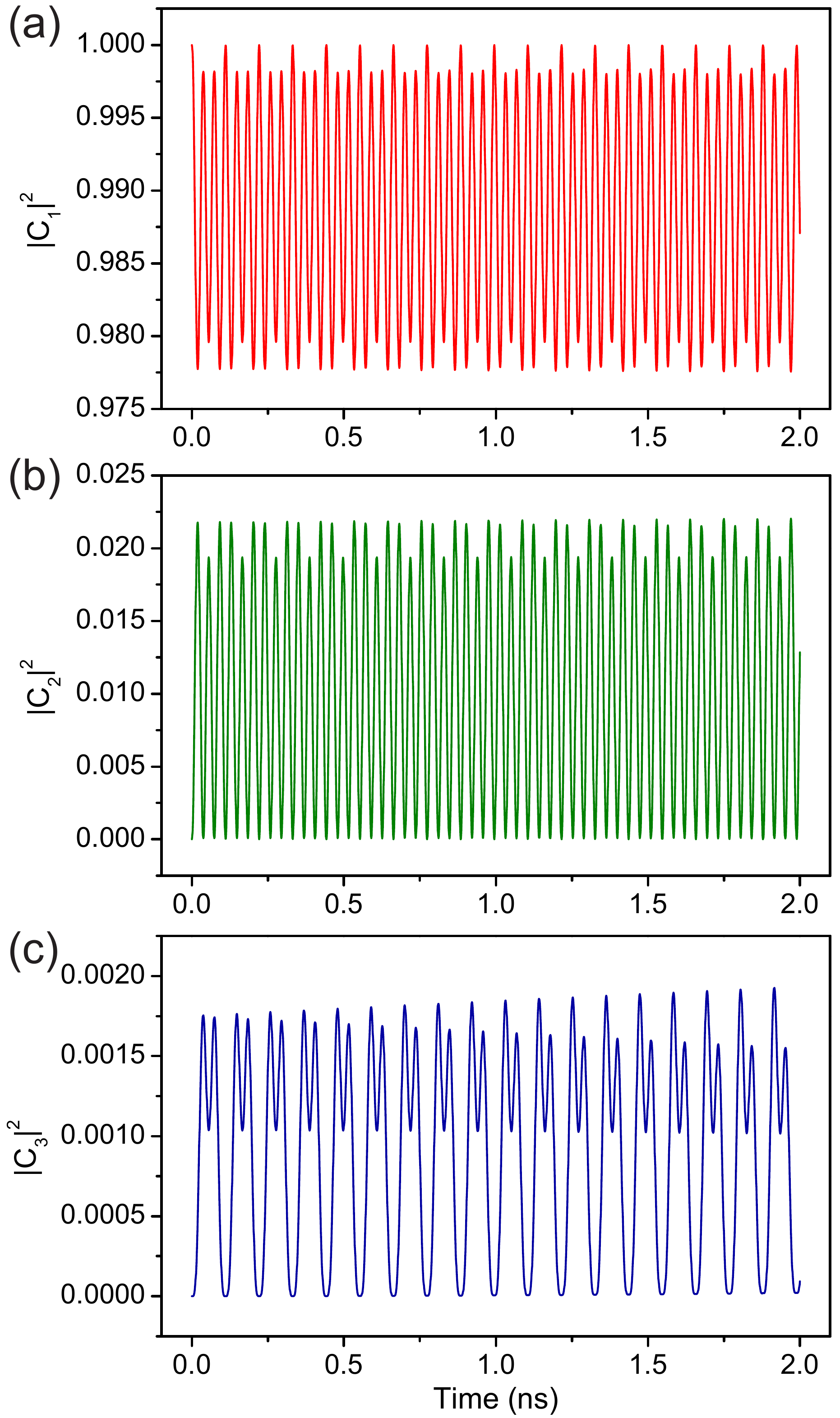}
\caption{\textbf{The population of energy levels.} For a straightforward clarification, we use a square-pulse model (i.e., the pulse intensity is a constant during the interval of 2 ns) to calculate the population coefficients $|C_1|^2$, $|C_2|^2$ and $|C_3|^2$. The main information we can get is the dominant population of atoms. We also can see a transient optical effect associated with rapid oscillations before the superposition states decay to equilibrium states.
} \label{Fig5}
\end{figure*}

\begin{figure*}
\includegraphics[width=0.85\textwidth]{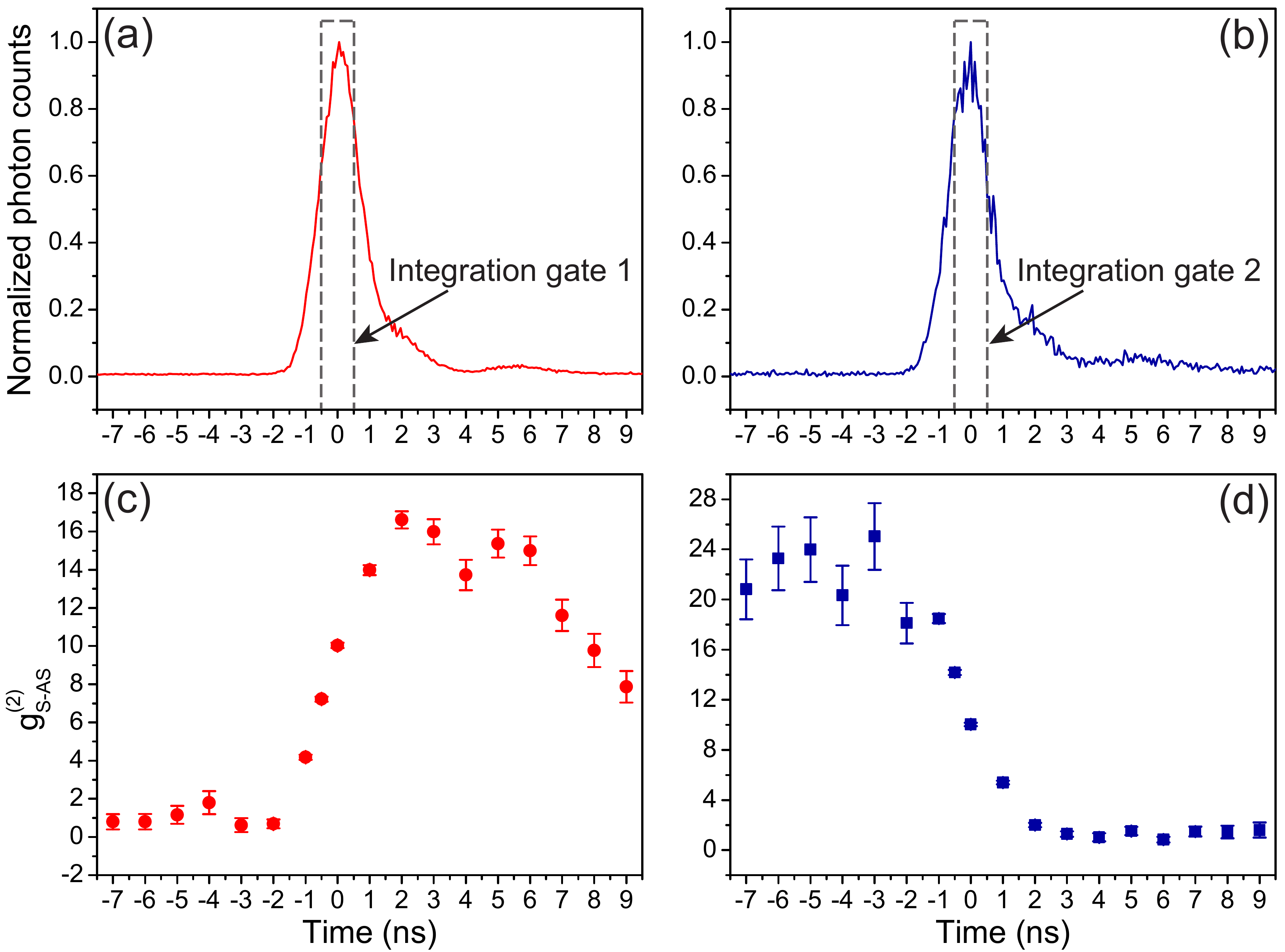}
\caption{\textbf{The time dependence of cross correlation.} \textbf{(a)} and {(b)} show the normalized photon counts of signal photons and idler photons respectively. By scanning the delay of the integration gate 2, we obtain \textbf{(c)} while the integration gate 1 is centered at 0 ns. By scanning the delay of the integration gate 1, we obtain \textbf{(d)} while the integration gate 2 is centered at 0 ns. The beam waist is 200\,$\mu {\rm m}$, and the pulse energy is 225\,pJ.}
\end {figure*}

%%%%%%%%%%%%%%%% added %%%%%%%%%%%%%%%%%%%%
Suppose that the eigenstates of the matrix
\begin{equation}\label{matrix5}
\frac{\hbar}{2}
\left(
{\begin{array}{*{20}c}
   { - 2\Delta _1 } & {\Omega _{12} } & {0}  \\
   {\Omega _{21} } & {2\Delta _2 } & {\Omega _{23} }  \\
   {0} & {\Omega _{32} } & { - 2\Delta _3 }  \\
\end{array}}
\right) 
\end{equation}
is 
\begin{equation} \label{eigenstate}
\left( {\begin{array}{*{20}c}
   {f_n }  \\
   {g_n }  \\
   {h_n }  \\
\end{array}} \right){\rm{exp}}( - i\lambda _n t)  \, ,
\end{equation}
where $n=1, 2, 3$ and
\begin{equation}
\left( {\begin{array}{*{20}c}
   {f_n }  \\
   {g_n }  \\
   {h_n }  \\
\end{array}} \right)
\end{equation}
is independent of time $t$. With the matrix (\ref{matrix5}), (\ref{eigenstate}) and a Schr\"{o}dinger equation, we can obtain the following equation.
\begin{equation}
\begin{array}{l}
\frac{\hbar }{2}\left( {\begin{array}{*{20}c}
   { - 2\Delta _1  - 2\lambda _n } & {\Omega _{12} } & {}  \\
   {\Omega _{21} } & {2\Delta _2  - 2\lambda _n } & {\Omega _{23} }  \\
   {} & {\Omega _{32} } & { - 2\Delta _3  - 2\lambda _n }  \\
\end{array}} \right)\\
\times \left( {\begin{array}{*{20}c}
   {f_n }  \\
   {g_n }  \\
   {h_n }  \\
\end{array}} \right){\rm{exp}}( - i\lambda _n t) = 0 \, ,
\end{array}
\end{equation}
Base on this equation, the eigenstates and eigenvalues can be calculated out.

Suppose $C_n  = a_n \exp \left [ (-1)^n i\Delta _n t \right ]$. The expansion coefficients $a_n$ can be expressed in terms of the eigenstates of the matrix (\ref{matrix5}). 

\begin{equation}
\begin{aligned}
\left( {\begin{array}{*{20}c}
   {a_1 }  \\
   {a_2 }  \\
   {a_3 }  \\
\end{array}} \right) &= \sum\limits_{n = 1}^3 {\left\{ {q_n \left( {\begin{array}{*{20}c}
   {f_n }  \\
   {g_n }  \\
   {h_n }  \\
\end{array}} \right)\exp ( - i\lambda _n t)} \right\}}  \\
& = \left( {\begin{array}{*{20}c}
   {G_{11} } & {G_{12} } & {G_{13} }  \\
   {G_{21} } & {G_{22} } & {G_{23} }  \\
   {G_{31} } & {G_{32} } & {G_{33} }  \\
\end{array}} \right)\left( {\begin{array}{*{20}c}
   {\exp ( - i\lambda _1 t)}  \\
   {\exp ( - i\lambda _2 t)}  \\
   {\exp ( - i\lambda _3 t)}  \\
\end{array}} \right)
\end{aligned}
\end{equation}
where $q_n$ and $G_{mn}$ is expansion coefficient. Note that the coefficients $C_1 =1, C_2=0, C_3=0$ at initial time ($t=0$). Then, we can calculate out the expansion coefficients $G_{mn}$ and obtain the wave function $\psi$. 
\begin{equation}
\begin{aligned}
 \psi  =& \sum\limits_{n = 1}^3 {\left[ {C_n (t)\left| n \right\rangle \exp ( - i\omega _n t)} \right]}  \\ 
 =& \left| 1 \right\rangle \{  G_{11} \exp \left[ { - i(\omega _1  + \Delta _1  + \lambda _1 )t} \right]+ \\
&\qquad G_{12} \exp \left[ { - i(\omega _1  + \Delta _1  + \lambda _2 )t} \right]+ \\
&\qquad G_{13} \exp \left[ { - i(\omega _1  + \Delta _1  + \lambda _3 )t} \right] \} \\ 
+&  \left| 2 \right\rangle \{ G_{21} \exp \left[ { - i(\omega _2  - \Delta _2  + \lambda _1 )t} \right]+\\
&\qquad G_{22} \exp \left[ { - i(\omega _2  - \Delta _2  + \lambda _2 )t} \right]+ \\
&\qquad G_{23} \exp \left[ { - i(\omega _2  - \Delta _2  + \lambda _3 )t} \right] \} \\ 
+&  \left| 3 \right\rangle \{ G_{31} \exp \left[ { - i(\omega _3  + \Delta _3  + \lambda _1 )t} \right]+ \\
&\qquad G_{32} \exp \left[ { - i(\omega _3  + \Delta _3  + \lambda _2 )t} \right]+ \\
&\qquad G_{33} \exp \left[ { - i(\omega _3  + \Delta _3  + \lambda _3 )t} \right] \} \\ 
 \end{aligned}
\end{equation}

Initially, atoms are prepared in state $\left| 1 \right\rangle $. However, during the light-matter interaction process, the population of each energy level will change, and the bare states will change to a coherent superposition state $\psi$ as is shown above. Figure\,5\,(a), (b) and (c) show the population of the energy level $\left| 1 \right\rangle $, $\left| 2 \right\rangle $ and $\left| 3 \right\rangle $ respectively. We can see that most atoms populate on the level $\left| 1 \right\rangle $. There is a rapid oscillation in the population, which is called the transient optical effect. If the decay for every energy level is considered, the rapid oscillation will behave damped and the superposition state $\psi$ will finally become an equilibrium state which is independent of time. In our experiment, the interaction time is so short (about 2 ns) that the decay effect is not taken into account.

%%%%%%%%%%%%%%%%%%%%% added %%%%%%%%%%%%%%%%%%%%%%%%%%%%%%%
The expectation value of the dipole moment is
\begin{equation}
\begin{array}{*{20}l}
\langle \psi \left | ex \right |  \psi \rangle &=&D_1 \exp\left[ -i\left( \omega _0 +\lambda_2-\lambda_1 \right)t\right ]\\
&+&D_2 \exp\left[ -i\left( \omega _0 +\lambda_3-\lambda_1 \right)t\right ]\\
&+&D_3 \exp\left[ -i\left( \omega _0 +\lambda_2-\lambda_3 \right)t\right ]\\
&+&D_4 \exp\left(-i \omega _0 t\right)\\
&+&D_5 \exp\left[ -i\left( \omega _0 +\lambda_3-\lambda_2 \right)t\right ]\\
&+&D_6 \exp\left[ -i\left( \omega _0 +\lambda_1-\lambda_3 \right)t\right ]\\
&+&D_7 \exp\left[ -i\left( \omega _0 +\lambda_1-\lambda_2 \right)t\right ]\\
&+& {\rm c.c.}
\end{array}
\end{equation}
where $x$ is the projection of $\vec R$ on the polarization direction of the coupling light, and $D_j$ denotes the relative amplitude of the corresponding spectral component.
\begin{equation}
\left( {\begin{array}{*{20}c}
   {D_1 }  \\
   {D_2 }  \\
   {D_3 }  \\
   {D_4 }  \\
   {D_5 }  \\
   {D_6 }  \\
   {D_7 }  \\
\end{array}} \right){\rm{ = }}\left( {\begin{array}{*{20}c}
   {d_{12} G_{11} G_{22}  + d_{32} G_{31} G_{22} }  \\
   {d_{12} G_{11} G_{23}  + d_{32} G_{31} G_{23} }  \\
   {d_{12} G_{13} G_{22}  + d_{32} G_{33} G_{22} }  \\
   \{{d_{12} \left[ {G_{11} G_{21}  + G_{12} G_{22}  + G_{13} G_{23} } \right] }\\
+ {d_{32} \left[ {G_{31} G_{21}  + G_{32} G_{22}  + G_{33} G_{23} } \right]} \} \\
   {d_{12} G_{12} G_{23}  + d_{32} G_{32} G_{23} }  \\
   {d_{12} G_{13} G_{21}  + d_{32} G_{33} G_{21} }  \\
   {d_{12} G_{12} G_{21}  + d_{32} G_{32} G_{21} }  \\
\end{array}} \right)
\end{equation}
where $d_{mn}= \left\langle m \right|ex\left| n \right\rangle $ is the effective dipole moment corresponds to the transtion $\left| n \right\rangle  \to \left| m \right\rangle $. So far, we have derived the frequency position and relative emission amplitude of each spectral component with a straightforward model. Initially, atoms are prepared in state $\left| 1 \right\rangle $. The photons stem from a same initial state, which is a necessary condition of generation of non-classically correlation. And not only that, there should also exists a bunching effect for the emissions of the signal and idler photons\cite{Nienhuis1993}.

In addition, during the light-matter interference, there exists AC-Stark shift of the energy level, i.e. the detuning $\Delta _n$ will change to $(-1)^n\lambda_n$. The AC-Stark shift of the state $ \left| n \right\rangle $ is equal to $\left [( - 1)^n  \times \lambda _n {\rm{ - }}\Delta _n \right]$ which is on the order of $2\pi \times 100$\,MHz in our experiment. It is very straightforward to calculate the AC-Stark shift of each state, which is the reason why we depict the detuning in the unusual way as is shown in Fig.\,1(a).

\subsection{\textbf {Measurement of time dependence of cross correlation:}}
The detection system always has a minimum resolving time and a minimum coincidence window (also known as integration gate). The results in Fig.\,4 are obtained by setting a 9-ns integration gates centered at the arrival time of the desired photons. In order to measure the time dependence of cross correlation, we employ a new counting system whose resolving time is shorter than 1 ns and the integration gate is 1 ns. 

The new experiment reveals more interesting dynamic properties of the light-matter interference. Fig.\,6 (a) and (b) show the normalized photon counts of signal photons and idler photons respectively. In Fig.\,6 (a) and (b), the delay of the two integration gates is 0 ns. By scanning the delay of the integration gate 2, we obtain Fig\,6 (c) while the integration gate 1 is centered at 0 ns. Similarly, we obtain Fig\,6 (d) by scanning the delay of the integration gate 1 while the integration gate 2 is centered at 0 ns. In the time region around 0 ns, there is a monotone zone of the cross correlation. A higher cross correlation can be obtained when the integration gate 1 is ahead of the integration gate 2. In contrast, the cross correlation is lower when the integration gate 2 is ahead of the integration gate 1. This monotone zone implies that the generation time of idler photons is posterior to that of the correlated signal photons. Thus, there is a cascaded-like emission of photon pairs. 

In our experiment, the switching extinction ratio of the coupling light is several hundreds, which means the coupling light is not absolutely zero when it should have been switched off. The non-zero coupling light may excite signal or idler photons in the time region far away from 0 ns, of course in a very low probability. We can see that the statistic error bar of the data far away from 0 ns is much larger than that around 0 ns, which means a very low excitation probability when either of the integration gates is set far away from 0 ns.

\end{document}